\documentclass[preprint,floatfix] {revtex4} 

\usepackage{graphicx}
\usepackage{subfigure}
\begin{document}

\title{Calculation of the bound states of power-law and logarithmic potentials
through a generalized pseudospectral method}
\author{Amlan K. Roy}
\affiliation{Department of Chemistry, University of New Brunswick, 
Fredericton, NB, E3B 6E2, Canada}
\email{akroy@unb.ca}

\begin{abstract}
Bound states of the power-law and logarithmic potentials are calculated using
a generalized pseudospectral method. The solution of the single-particle 
Schr\"odinger equation in a nonuniform and optimal spatial discretization
offers accurate eigenvalues, densities and expectation values. The calculations
are carried out for states with arbitrary $n$ and $\ell$ quantum numbers. 
Comparisons are made with the available literature data and excellent agreement
is observed. In all  
the cases, the present method yields considerably improved results over the 
other existing calculations. Some new states are reported. 
\end{abstract}
\maketitle

\section{Introduction}
Quantum mechanical treatment of a large number of important physical processes
in various branches of physics and chemistry often requires solving the 
time-independent Schr\"odinger equation involving a central potential. For 
example, the Hellmann potential [1] has been used to study the electron-core or 
electron-ion interactions or the atomic inner-shell ionization problems, the 
short-range Hulth\'en potential [2] is relevant in the nuclear and particle 
physics, the exponential cosine screened Coulomb potential and the celebrated 
Morse potential [3] found 
applications in solid-state physics, etc. Unfortunately, the exact 
solutions of such systems are obtainable only in an exceptionally few 
instances and for almost all practical purposes, one has to take resort to the
approximation methods or direct numerical techniques. Therefore, almost ever 
since the inception of quantum mechanics, a large number of attractive 
formalisms have been proposed by many workers to calculate the eigenvalues
and eigenfunctions of the spherically symmetric potentials, varying in terms of
complexity, efficiency and elegancy. Significant strides have been made over 
the years and it is still an active area of research. Possibly the most widely 
used approximation schemes are the Rayleigh-Schr\"odinger perturbation theory 
and the Rayleigh-Ritz variational method. Both of these methods have witnessed 
outstanding successes to deliver physically meaningful results for many 
systems of interest. The other popular schemes are the 1/N expansion or many 
of its variants, the WKB method, etc. However the important shortcomings in 
some of these methods are that these often involve  extensive and elaborate 
algebraic manipulations and the limited nature of the eigenvalues and 
eigenfunctions to be expressible in compact analytical forms. 

This work is devoted to a detailed calculation of the power-law and the 
logarithmic potentials which have relevant applications in the field of 
particle physics 
[4-13]. These potentials have been studied from various perspectives by
several researchers employing a 
number of approximations; e.g., the WKB treatment [14], the shifted 1/N 
expansion method [15-17], the variational technique [18], through an 
interpolation formula [19] and also by the direct numerical integration methods
[14,19]. Although several formally attractive and elegant formalisms exist in 
the literature, there is lack of accurate eigenvalues, especially those states
characterized by higher quantum numbers and the eigenfunctions. Here we employ a generalized pseudospectral (GPS) method for this purpose 
which has shown considerable promise in the field of static and dynamic processes
in atomic and molecular systems recently (see, e.g., [20-23]) involving mainly
the Coulomb potentials. However, to our knowledge, no attempts have been made 
so far to apply this scheme 
to other potentials of physical interest including the ones of current study. 
Therefore the purpose of this article is two-fold, (a) To assess the validity
and performance of the GPS method for the power-law and logarithmic 
potentials, and (b) To determine accurately the bound state spectra of these 
systems. As is demonstrated in Section 3, it appears that the current 
method is capable of producing excellent results of these systems; for the low
as well as  
higher states. The layout of the article is as follows. Section 2 presents 
a brief overview of the method of calculation. In Section 3 we first make 
some test calculations on some simple quantum mechanical systems and subsequently present the results for the power-law 
and the logarithmic potentials. Finally we make a few concluding remarks in 
Section 4. 

\section{Methodology}\label{sec:method}
In this section, we present an overview of the generalized pseudospectral 
method (GPS) employed to solve the radial eigenvalue problem with the power-law
and the logarithmic potentials. A more detailed account can be found in the refs. 
[24,25,20-23].

The desired radial Schr\"odinger equation to be solved, can be written in the 
following operator form, 
\begin{equation}
\hat{H}(r)\ \phi(r) =\varepsilon \ \phi(r),
\end{equation}
where the Hamiltonian includes the usual kinetic and potential energy 
operators,  
\begin{equation}
\hat{H}(r) =-\frac{1}{2} \ \ \frac{d^2}{dr^2} +v(r),
\end{equation}
with 
\begin{equation}
 v(r) = V(r) + \frac{\ell (\ell+1)}{2r^2} 
\end{equation}
and 
$V(r)=\mathrm{sgn}(\nu) Ar^{\nu}(\nu>-2)$ or $V(r)=\ln r $. The symbols 
have their usual significances. 

The majority of the finite-difference discretization schemes for the solution
of the radial Schr\"odinger equation arising in these situations often require
a large number of spatial grid points predominantly because of the uniform 
distributional nature of the spatial grid. The GPS method, on the other hand 
allows the nonuniform and optimal spatial discretization, and as is mentioned 
below, maintains the same accuracy at both the small and large $r$ regions. 
Therefore one has the advantage of working with a much lesser grid points compared
to many other methods in the literature, and can have a denser mesh at the smaller
$r$ while a coarser mesh at the large $r$. 
The principal feature of this scheme is that a function $f(x)$ defined in 
the interval $x \in [-1,1]$ can be approximated by the polynomial $f_N(x)$ 
of order N so that,  
\begin{equation}
f(x) \cong f_N(x) = \sum_{j=0}^{N} f(x_j)\ g_j(x),
\end{equation}
and the approximation is \emph {exact} at the \emph {collocation points} 
$x_j$,i.e.,
\begin{equation}
f_N(x_j) = f(x_j).
\end{equation}
In the Legendre pseudospectral method that we use in this study,  
$x_0=-1$, $x_N=1$, and the $x_j (j=1,\ldots,N-1)$ are obtained from the roots 
of the first derivative of the Legendre polynomial $P_N(x)$ with respect to 
$x$, i.e., 
\begin{equation}
P'_N(x_j) = 0.
\end{equation}
The $g_j(x)$ in Eq.~(2) are called the cardinal functions and given by the 
following expression,
\begin{equation}
g_j(x) = -\frac{1}{N(N+1)P_N(x_j)}\ \  \frac{(1-x^2)\ P'_N(x)}{x-x_j},
\end{equation}
satisfying the unique property $g_j(x_{j'}) = \delta_{j'j}$.
Now the semi-infinite domain $r \in [0, \infty]$ is mapped into the finite 
domain $x \in [-1,1]$ by the transformation $r=r(x)$. Now the following 
algebraic nonlinear mapping is introduced
\begin{equation}
r=r(x)=L\ \ \frac{1+x}{1-x+\alpha},
\end{equation}
where L and $\alpha=2L/r_{max}$ may be termed as the mapping parameters. At this
stage introduction of the following relation, 
\begin{equation}
\psi(r(x))=\sqrt{r'(x)} f(x)
\end{equation}
coupled with the symmetrization procedure [31,32] leads to the transformed 
Hamiltonian as below, 
\begin{equation}
\hat{H}(x)= -\frac{1}{2} \ \frac{1}{r'(x)}\ \frac{d^2}{dx^2} \ \frac{1}{r'(x)}
+ v(r(x))+v_m(x),
\end{equation}
where $v_m(x)$ is given by the following relation,
\begin{equation}
v_m(x)=\frac {3(r'')^2-2r'''r'}{8(r')^4}.
\end{equation}
The advantage is that this leads to a \emph {symmetric} matrix eigenvalue 
problem which can be readily solved to give accurate eigenvalues and 
eigenfunctions and at the same time. For the particular transformation 
used above, $v_m(x)=0$. This discretization then finally leads to the following
set of coupled equations, 
\begin{widetext}
\begin{equation}
\sum_{j=0}^N \left[ -\frac{1}{2} D^{(2)}_{j'j} + \delta_{j'j} \ v(r(x_j))
+\delta_{j'j}\ v_m(r(x_j))\right] A_j = EA_{j'},\ \ \ \ j=1,\ldots,N-1,
\end{equation}
\end{widetext}
where
\begin{equation}
A_j  = \left[ r'(x_j)\right]^{1/2} \psi(r(x_j))\ \left[ P_N(x_j)\right]^{-1}.
\end{equation}
and the symmetrized second derivative of the cardinal function, $D^{(2)}_{j'j}$
is given by,
\begin{equation}
D^{(2)}_{j'j} =  \left[r'(x_{j'}) \right]^{-1} d^{(2)}_{j'j} 
\left[r'(x_j)\right]^{-1}, 
\end{equation}
with
\begin{eqnarray}
d^{(2)}_{j',j} & = & \frac{1}{r'(x)} \ \frac{(N+1)(N+2)} {6(1-x_j)^2} \ 
\frac{1}{r'(x)}, \ \ \ j=j', \nonumber \\
 & & \nonumber \\
& = & \frac{1}{r'(x_{j'})} \ \ \frac{1}{(x_j-x_{j'})^2} \ \frac{1}{r'(x_j)}, 
\ \ \ j\neq j'.
\end{eqnarray}
In order to make a judicious choice of the mapping parameters, a large number of 
tests have been made to check the performance of this scheme for a broad range 
of parameter sets in these potentials available in the literature. The results 
are reported only up to the precision that were found to maintain stability with
respect to these variations. In this way, a consistent set of the numerical parameters 
($r_{max}=200,$ $\alpha=25$ and $N=300$) has been chosen which seemed to be 
appropriate and satisfactory for the current problem. 

\section{Results and Discussion}
First we present results for two simple test cases where accurate results are 
available for comparison. The first one is the well-known Morse potential [26], having the 
following form,
\begin{equation}
V(r)=25(e^{-4(r-3)}-2e^{-2(r-3)})
\end{equation}
The above potential supports four bound states and the corresponding energies 
are given by,
\begin{equation}
E_n=-[5-\sqrt{2}(n+\frac{1}{2})]^2
\end{equation}
where $n=0,1,2,3.$ Table I presents the calculated eigenvalues along with the
exact analytical values [26]and the B-spline results [27]. While for the 
ground and second excited states, the current results match the exact values 
completely, slight overestimation is noticed for the first and third excited 
states. It may be noted that the first three states reached the precision of 
Table I with an $r_{max}$ of 50 a. u. and with 200 grid points. However, the 
fourth state required a grid of size 200 a. u. having 300 radial points, 
presumably due to its weakly bound nature and therefore having a long tail. 
The B-spline basis set calculations were done using a box of radius 250 a. u. 
and 263 B-splines.  

\begingroup
\squeezetable
\begin{table}
\caption {\label{tab:table1}Comparison of the eigenvalues E (in a. u.) of the 
Morse potential with the literature data.} 
\begin{ruledtabular}
\begin{tabular}{cccc}
 $n$ & \multicolumn{3}{c}{Energy}   \\ 
\cline{2-4}
     & This work    & B-splines (Ref. [27]) &  Exact (Ref. [26])   \\
\hline
 0 & $-18.428932188134$  & $-18.428932188135$ & $-18.428932188134$  \\
 1 & $-8.2867965644036$  & $-8.286796564404$  & $-8.2867965644035$  \\
 2 & $-2.1446609406726$  & $-2.144660940673$  & $-2.1446609406726$  \\
 3 & $-0.0025253169419$  & $-0.002525316942$  & $-0.0025253169416$  \\
\end{tabular}
\end{ruledtabular}
\end{table}
\endgroup

As a second case we take the common and frequently studied anharmonic 
oscillator potential having a quartic perturbation given as,
\begin{equation}
V(r)=\frac{1}{2}m \omega^2r^2 + \frac{1}{4}\lambda r^4, \ \ \ \lambda > 0
\end{equation}
This potential has been studied quite extensively in the literature as a 
testing ground for many theoretical methodologies. Table II shows the 
agreement of the current results with the numerically calculated ``exact'' 
values of [28] for selected values of the quantum numbers. A very large segment
of the quantum numbers are chosen covering ground to the very high excited 
states (having high values for both the radial as well as the angular quantum 
numbers), and for all these states the GPS results reproduce the reference values 
nicely. A careful glance through the table shows that within a given $\ell$, the 
eigenvalues show better accuracy as the quantum number $n$ increases. Such
features have been observed in other works also (e.g., the phase-integral 
calculations [28]).

\begingroup
\squeezetable
\begin{table}
\caption {\label{tab:table2}The calculated eigenvalues E (in a. u.) of the 
anharmonic oscillator with $m=1,\ \omega=1,\ \lambda=2$ for the various 
combinations of the quantum numbers $n$ and $\ell$. The 
numerically calculated ``exact'' reference values are quoted from [28].} 
\begin{ruledtabular}
\begin{tabular}{cccccccc}
$n$    & $\ell$ & \multicolumn{2}{c}{Energy} & $n$ & $\ell$ & 
\multicolumn{2}{c}{Energy} \\ 
\cline{3-4} \cline{7-8}
  &  & This work & Exact (Ref. [28]) &  &  & This work & Exact (Ref. [28]) \\
\hline
 0 & 0 &   2.32440635210  &   2.324406352106 &  5 & 3  &  
13.94920888000  &  13.949208880004 \\
 2 & 0 &   6.57840194902  &   6.578401949025 & 10 & 4  &  
29.51001260726  &  29.510012607268 \\
10 & 0 &  30.06476147957  &  30.064761479579 & 5  & 5  &  
13.26445877906  &  13.264458779062 \\
50 & 0 & 213.9909056603   & 213.990905660342 & 10 & 10 &  
27.09249230522  &  27.092492305227 \\
 1 & 1 &   4.19017126505  &   4.190171265052 & 50 & 20 & 
209.4822651210  & 209.482265121044 \\
 5 & 1 &  14.33886608777  &  14.338866087774 & 50 & 50 & 
187.5297080140  & 187.529708014003 \\  
 2 & 2 &   6.24277802550  &   6.242778025500 &    &    &    
                &                  \\
\end{tabular}
\end{ruledtabular}
\end{table}
\endgroup

After these tests we now in Table III present the ground state energies (in units
of $\hbar=2m=1$) for the various power-law potentials having several values of $\nu$. 
The first row shows the results for $\nu=-1$ and 2 corresponding to the well-known 
Coulomb and the harmonic oscillator potentials for which exact solutions are known.  
In both these cases, our calculation reproduces the exact results very well. The 
variational as well as the numerical results exist for all of these states
excepting $\nu=2.5$ and 3.5 
For the $\nu=4$ case, several accurate calculations are available in the literature
including the large-order shifted 1/N expansion [17] and the present result is in 
complete agreement with the most accurate ``exact'' 
result of [17]. However to the best of our knowledge, such definitive and accurate 
results are not available for the other members of this series and the present values
show considerable improvement over the existing results. The GPS results are more in 
conformity with the numerical results of [16] but noticeably deviate from the values
of [18].

\begingroup
\squeezetable
\begin{table}
\caption {\label{tab:table3}The calculated ground states ($n=0,\ \ell=0$) along
with the literature results for various power law potentials in units of
$\hbar=2m=1$.} 
\begin{ruledtabular}
\begin{tabular}{cccccc}
 $V(r)$ & \multicolumn{2}{c}{Energy}& $V(r)$ & \multicolumn{2}{c}{Energy} \\ 
\cline{2-3} \cline{5-6}
     & This work    & Literature  &       & This work     &  Literature   \\
\hline
$-r^{-1}$ &$-0.2500000000$ & $-0.25$\footnotemark[1], $-0.25$\footnotemark[2] &
$r^{2}$   & 3.000000000 & 3\footnotemark[1], 3.0\footnotemark[2]  \\
$r^{0.15}$ & 1.327945844 & 1.32798\footnotemark[1], 1.32795\footnotemark[2] &
$r^{2.5}$  & 3.242232312 &                 \\
$r^{0.5}$  & 1.833393609 & 1.83352\footnotemark[1], 1.83339\footnotemark[2] &
$r^{3.0}$  & 3.450562689 & 3.45110\footnotemark[1], 3.45056\footnotemark[2] \\
$r^{0.75}$ & 2.108136609 & 2.10829\footnotemark[1], 2.10814\footnotemark[2] &  
$r^{3.5}$  & 3.634394905 &                 \\
$r^{1.0}$  & 2.338107410 & 2.33825\footnotemark[1], 2.33810\footnotemark[3] &
$r^{4.0}$  & 3.799673029 & 3.80241\footnotemark[1], 3.79967\footnotemark[2] \\
$r^{1.5}$  & 2.708092416 & 2.70816\footnotemark[1], 2.70809\footnotemark[2] & 
           &   & 3.799673030\footnotemark[4], 3.799673029\footnotemark[5]   \\
\end{tabular}
\end{ruledtabular}
\footnotetext[1]{Ref. [18]}
\footnotetext[2]{Numerical results, ref. [16]}
\footnotetext[3]{Ref. [29]}
\footnotetext[4]{Ref. [17]}
\footnotetext[5]{Exact value, as quoted in ref. [17]}
\end{table}
\endgroup

Now we show the efficacy of the current method for the excited states by
presenting the first six eigenvalues corresponding to $\ell=0,1,\cdots,5$ 
in Table IV for the $r^{0.5}$ potential. The variational results are from [16],
while the numerical reference values are taken from [19] for the 
$n, \ell \leq 4$ states. As in the previous table, there is good agreement 
between the current and the numerical values. However, the results of [18] are 
consistently overestimated in all cases except the third and fourth states 
belonging to $\ell=0$. No results could be found for $\ell >4$. As in the last
table, here also we have produced superior results for these systems than the 
previous works in the literature. Some new states are reported. Additionally 
Table VI displays the energies 
(in a. u.) for the two potentials $V(r)=-2^{1.7}r^{-0.2}$ and $V(r)=2^{7/2}r$ 
which have been examined earlier by some workers.  All the 
states with $n \leq 4$ and $\ell \leq 3$ are calculated and compared wherever
possible. In the former case, accurate results are not available for comparison
and the present results match closely to all of them, while for the 
latter case, the GPS results are in excellent agreement with the accurate 
calculations [17]. 

\begingroup
\squeezetable
\begin{table}
\caption {\label{tab:table4}The first six eigenvalues E (in a. u.) of the 
$r^{0.5}$ potential along with the literature data for $\ell=0,1, \cdots, 5.$} 
\begin{ruledtabular}
\begin{tabular}{cccccc}
$\ell$ & \multicolumn{2}{c}{Energy}& $\ell$ & \multicolumn{2}{c}
{Energy} \\ 
\cline{2-3} \cline{5-6}
  & This work   & Literature  &  &  This work & Literature \\
\hline
0 & 1.83339360 & 1.83352\footnotemark[1], 1.83339\footnotemark[2] & 
3 & 2.95445093 & 2.95448\footnotemark[1], 2.95445\footnotemark[2] \\
  & 2.55064749 & 2.55152\footnotemark[1], 2.55065\footnotemark[2] &   
  & 3.35759134 & 3.35764\footnotemark[1], 3.35759\footnotemark[2] \\  
  & 3.05118194 & 3.05177\footnotemark[1], 3.05118\footnotemark[2] &
  & 3.70270499 & 3.70299\footnotemark[1], 3.70270\footnotemark[2] \\
  & 3.45213194 & 3.45197\footnotemark[1], 3.45213\footnotemark[2] &
  & 4.00736733 & 4.00796\footnotemark[1], 4.00737\footnotemark[2] \\
  & 3.79336044 & 3.79233\footnotemark[1], 3.79336\footnotemark[2] &
  & 4.28195944 & 4.28282\footnotemark[1], 4.28196\footnotemark[2] \\
  & 4.09392584 &                    &   & 4.53316865 &            \\
1 & 2.30049623 & 2.30056\footnotemark[1], 2.30050\footnotemark[2] &
4 & 3.21233437 & 3.21236\footnotemark[1], 3.21233\footnotemark[2] \\
  & 2.85433592 & 2.85473\footnotemark[1], 2.85434\footnotemark[2] &
  & 3.57275267 & 3.57275\footnotemark[1], 3.57275\footnotemark[2] \\
  & 3.28583329 & 3.28666\footnotemark[1], 3.28583\footnotemark[2] &
  & 3.88897564 & 3.88913\footnotemark[1], 3.88898\footnotemark[2] \\
  & 3.64738542 & 3.64838\footnotemark[1], 3.64739\footnotemark[2] &
  & 4.17268190 & 4.17308\footnotemark[1], 4.17268\footnotemark[2] \\
  & 3.96267650 & 3.96361\footnotemark[1], 3.96268\footnotemark[2] &
  & 4.43130627 & 4.43196\footnotemark[1], 4.43131\footnotemark[2] \\
  & 4.24465838 &                    &   & 4.66989741 &            \\
2 & 2.65756336 & 2.65760\footnotemark[1], 2.65756\footnotemark[2] & 
5 & 3.44244561 &                 \\
  & 3.12032849 & 3.12048\footnotemark[1], 3.12033\footnotemark[2] &
  & 3.77041929 &                 \\
  & 3.50245154 & 3.50296\footnotemark[1], 3.50245\footnotemark[2] &
  & 4.06336036 &                 \\
  & 3.83254391 & 3.83338\footnotemark[1], 3.83254\footnotemark[2] &
  & 4.32947933 &                 \\
  & 4.12580907 & 4.12686\footnotemark[1], 4.12581\footnotemark[2] &
  & 4.57430430 &                 \\
  & 4.39138573 &                    &   & 4.80174799 &            \\
\end{tabular}
\end{ruledtabular}
\footnotetext[1]{Ref. [18]}
\footnotetext[2]{Numerical results, ref. [19]}
\end{table}
\endgroup

\begingroup
\squeezetable
\begin{table}
\caption {\label{tab:table5}The calculated eigenvalues E (in a.u.) of the 
power-law potentials for several values of $n$ and $\ell.$ The left and the 
right hand sides correspond to the potentials $V(r)=-2^{1.7}r^{-0.2}$ and 
$V(r)=2^{7/2}r$ respectively.} 
\begin{ruledtabular}
\begin{tabular}{cccccc}
 \multicolumn{2}{c}{Energy}& $n$ & $\ell$ & \multicolumn{2}{c}
{Energy} \\ 
\cline{1-2} \cline{5-6}
 This work   & Literature  &  &  &  This work & Literature \\
\hline
$-2.68602822$ & $-2.68601$\footnotemark[1], $-2.6859$\footnotemark[2],
$-2.686$\footnotemark[3] & 0 & 0 & 9.352429641 &
9.352429643\footnotemark[4], 9.3524296418\footnotemark[5] \\
$-2.25351412$ & $-2.25483$\footnotemark[1], $-2.2530$\footnotemark[2],
$-2.253$\footnotemark[3] & 1 & 0 & 16.35179777 & 
16.35179777\footnotemark[4], 16.3517977765\footnotemark[5] \\  
$-2.04431800$ & $-2.04658$\footnotemark[1], $-2.0440$\footnotemark[2],
$-2.044$\footnotemark[3] & 2 & 0 & 22.08223931 & 
22.08223931\footnotemark[4], 22.0822393124\footnotemark[5] \\
$-1.91063527$ &     & 3 & 0 & 27.14683236 &                \\
$-1.81414352$ &     & 4 & 0 & 31.77653434 &                \\
$-2.34494617$ & $-2.34494$\footnotemark[1], $-2.3449$\footnotemark[2],
$-2.345$\footnotemark[3] & 0 & 1 & 13.44501809 &           \\
$-2.10073849$ & $-2.10103$\footnotemark[1], $-2.1006$\footnotemark[2],
$-2.101$\footnotemark[3] & 1 & 1 & 19.53780737 &           \\
$-1.95072177$ & $-1.95147$\footnotemark[1], $-1.9504$\footnotemark[2],
$-1.951$\footnotemark[3] & 2 & 1 & 24.83049317 &           \\
$-1.84490090$ &          & 3 & 1 & 29.62266174 &           \\
$-1.76427587$ &          & 4 & 1 & 34.06093721 &           \\
$-2.15626090$ & $-2.15626$\footnotemark[1], $-2.1562$\footnotemark[2],
$-2.156$\footnotemark[3] & 0 & 2 & 16.99272902 &           \\
$-1.99005560$ & $-1.99015$\footnotemark[1], $-1.9900$\footnotemark[2],
$-1.990$\footnotemark[3] & 1 & 2 & 22.51883350 &           \\
$-1.87503225$ & $-1.87535$\footnotemark[1], $-1.8749$\footnotemark[2],
$-1.875$\footnotemark[3] & 2 & 2 & 27.47553075 &           \\
$-1.78852162$ &          & 3 & 2 & 32.03881169 &           \\
$-1.71993045$ &          & 4 & 2 & 36.30801220 &           \\
$-2.02906490$ & $-2.02906$\footnotemark[1], $-2.0291$\footnotemark[2],
$-2.029$\footnotemark[3] & 0 & 3 & 20.20370253 &           \\
$-1.90486674$ & $-1.90491$\footnotemark[1], $-1.9049$\footnotemark[2],
$-1.905$\footnotemark[3] & 1 & 3 & 25.32846149 &           \\
$-1.81250205$ & $-1.81266$\footnotemark[1], $-1.8124$\footnotemark[2]
                         & 2 & 3 & 30.01858256 &           \\
$-1.73987512$ &          & 3 & 3 & 34.38846804 &           \\
$-1.68053730$ &          & 4 & 3 & 38.50906805 &           \\
\end{tabular}
\end{ruledtabular}
\footnotetext[1]{Ref. [16]}
\footnotetext[2]{Ref. [18]}
\footnotetext[3]{Numerical results, from [16]}
\footnotetext[4]{Ref. [17]}
\footnotetext[5]{Exact values, as quoted in [17]}
\end{table}
\endgroup

Table VI displays the computed eigenvalues for selected $n$ and $\ell$ quantum
numbers of the logarithmic potential. The numerical results [4] exist for all 
the states with $n,\ \ell \leq 4$, while for states with $n \leq 4, \ell \leq 3$ 
and $\ell=4;\ n=3,4$, the shifted 1/N expansion [16] as well as the variational 
results [16] are also available. Results are presented for $n$ and $\ell$ ranging
up to a maximum of 7 and 10. There is a scarcity of accurate reference values for
the comparison of these states and the present results show general agreement with
them. Some of the states are reported here for the first time and may constitute
a useful reference for future purposes.  

\begingroup
\squeezetable
\begin{table}
\caption {\label{tab:table6}The calculated eigenvalues E (in a.u.) of the 
logarithmic potential for several values of $n$ and $\ell.$} 
\begin{ruledtabular}
\begin{tabular}{cccccccc}
$n$ & $\ell$ & \multicolumn{2}{c}{Energy}& $n$ & $\ell$ & \multicolumn{2}{c}
{Energy} \\ 
\cline{3-4} \cline{7-8}
  &   & This work   & Literature  &  &  &  This work & Literature \\
\hline
0 & 0 & 1.04433226 & 
        1.04436\footnotemark[1],1.0445\footnotemark[2],1.0443\footnotemark[3] & 
0 & 3 & 2.28414135 & 
        2.28414\footnotemark[1],2.2842\footnotemark[2],2.286\footnotemark[3]  \\
1 & 0 & 1.84744258 & 
        1.84457\footnotemark[1],1.8485\footnotemark[2],1.8474\footnotemark[3]  & 
1 & 3 & 2.57978331 & 
        2.57967\footnotemark[1],2.5798\footnotemark[2],2.581\footnotemark[3]  \\  
2 & 0 & 2.28961571 & 
        2.28417\footnotemark[1],2.2903\footnotemark[2],2.2897\footnotemark[3]  & 
2 & 3 & 2.81044538 & 
        2.80999\footnotemark[1],2.8106\footnotemark[2],2.811\footnotemark[3]  \\
3 & 0 & 2.59570686 & 
        2.58863\footnotemark[1],2.5957\footnotemark[2],2.5957\footnotemark[3]  & 
3 & 3 & 2.99916581 & 
        2.99822\footnotemark[1],2.9996\footnotemark[2],2.999\footnotemark[3]  \\
4 & 0 & 2.82992843 & 
        2.82176\footnotemark[1],2.8293\footnotemark[2],2.8299\footnotemark[3]  & 
4 & 3 & 3.15866751 & 
        3.15719\footnotemark[1],3.1592\footnotemark[2],3.159\footnotemark[3]  \\
5 & 0 & 3.01965502 &                        & 5 & 3 & 3.29668751 &            \\
6 & 0 & 3.17910756 & 
        3.16956\footnotemark[1],3.17911\footnotemark[3]                        & 
0 & 4 & 2.49711469 & 2.499\footnotemark[3]                                    \\
7 & 0 & 3.31662376 &                        & 
1 & 4 & 2.74154358 & 2.742\footnotemark[3]                                    \\
0 & 1 & 1.64114133 & 
        1.64114\footnotemark[1],1.6412\footnotemark[2],1.643\footnotemark[3]   & 
2 & 4 & 2.94004751 & 2.941\footnotemark[3]                                    \\
1 & 1 & 2.15094678 & 
        2.15023\footnotemark[1],2.1513\footnotemark[2],2.151\footnotemark[3]   & 
3 & 4 & 3.10686428 & 
        3.10629\footnotemark[1],3.1071\footnotemark[2],3.107\footnotemark[3]  \\
2 & 1 & 2.49094221 & 
        2.48897\footnotemark[1],2.4917\footnotemark[2],2.491\footnotemark[3]   & 
4 & 4 & 3.25056363 & 
        3.24960\footnotemark[1],3.2512\footnotemark[2],3.251\footnotemark[3]  \\
3 & 1 & 2.74559643 & 
        2.74244\footnotemark[1],2.7465\footnotemark[2],2.744\footnotemark[3]   & 
0 & 5 & 2.67263174 &                                                          \\
4 & 1 & 2.94900787 & 
        2.94484\footnotemark[1],2.9498\footnotemark[2],2.948\footnotemark[3]   & 
1 & 5 & 2.88099141 &                                                          \\
5 & 1 & 3.11827840 &                        & 0 & 6 & 2.82191040 &            \\
6 & 1 & 3.26318814 &                        & 1 & 6 & 3.00348669 &            \\
7 & 1 & 3.38984841 &                        & 0 & 7 & 2.95178152 &            \\
0 & 2 & 2.01330864 & 
        2.01331\footnotemark[1],2.0134\footnotemark[2],2.015\footnotemark[3]   & 
1 & 7 & 3.11268074 &                                                          \\
1 & 2 & 2.38743285 & 
        2.38718\footnotemark[1],2.3875\footnotemark[2],2.388\footnotemark[3]   & 
0 & 8 & 3.06671400 &                                                          \\
2 & 2 & 2.66249204 & 
        2.66160\footnotemark[1],2.6629\footnotemark[2],2.663\footnotemark[3]   & 
1 & 8 & 3.21116668 &                                                          \\
3 & 2 & 2.87949935 & 
        2.87786\footnotemark[1],2.8801\footnotemark[2],2.880\footnotemark[3]   & 
0 & 9 & 3.16979180 &                                                          \\
4 & 2 & 3.05848949 & 
        3.05610\footnotemark[1],3.0592\footnotemark[2],3.060\footnotemark[3]   & 
1 & 9 & 3.30084996 &                                                          \\
5 & 2 & 3.21070014 &                        & 0 & 10& 3.26323280 &            \\

\end{tabular}
\end{ruledtabular}
\footnotetext[1]{Ref. [16]}
\footnotetext[2]{Ref. [18]}
\footnotetext[3]{Numerical results, from [4]}
\end{table}
\endgroup

The usefulness of the method is further illustrated by calculating the wave functions
and the expectation values $\langle r^{-1} \rangle$ and $ \langle r^1 \rangle$ for 
both the potentials. Table VII shows a cross-section of the expectation values, for 
which no literature results could be found. The first three states belonging
to $\ell=0,1,2$ are 
given. Finally Figs. 1 and 2 depict the radial densities for the first four 
states of $\ell=0,1,2$ of the logarithmic and the $r^{0.5}$ case, along with the 
respective potentials. As expected they show the desired numbers of peaks. 

\begingroup
\squeezetable
\begin{table}
\caption {\label{tab:table7}The calculated expectation values (in a.u.) of the 
power-law and logarithmic potentials for the first three states belonging to
$\ell=0,1,2.$} 
\begin{ruledtabular}
\begin{tabular}{cccccccc}
$V(r)$ & $\ell$ & $\langle r^{-1} \rangle$ & $ \langle r^1 \rangle$ & $V(r)$  & $\ell$ & 
$ \langle r^{-1}\rangle$ & $ \langle r^1 \rangle $ \\
\hline
$\ln r$ & 0 & 0.975829609 & 1.39052517  & $r^{0.5}$ & 0 & 0.767168993  & 1.72566470 \\
        & 0 & 0.497961528 & 3.15106068  &           & 0 & 0.469136231  & 3.36898957 \\  
        & 0 & 0.339365144 & 4.91871111  &           & 0 & 0.354202831  & 4.82937427 \\
        & 1 & 0.493205837 & 2.38769029  &           & 1 & 0.437206279  & 2.65352685 \\
        & 1 & 0.327683049 & 4.14985349  &           & 1 & 0.323564598  & 4.16612795 \\
        & 1 & 0.247106879 & 5.91466141  &           & 1 & 0.263022498  & 5.55753612 \\
        & 2 & 0.330196264 & 3.38653900  &           & 2 & 0.315487133  & 3.50669405 \\
        & 2 & 0.245961257 & 5.14983438  &           & 2 & 0.253812312  & 4.93344373 \\
        & 2 & 0.196769673 & 6.91391694  &           & 2 & 0.215382759  & 6.27078423 \\

\end{tabular}
\end{ruledtabular}
\end{table}
\endgroup

\begin{figure}
\begin{minipage}[b]{0.40\textwidth}
\centering
\includegraphics[scale=0.27]{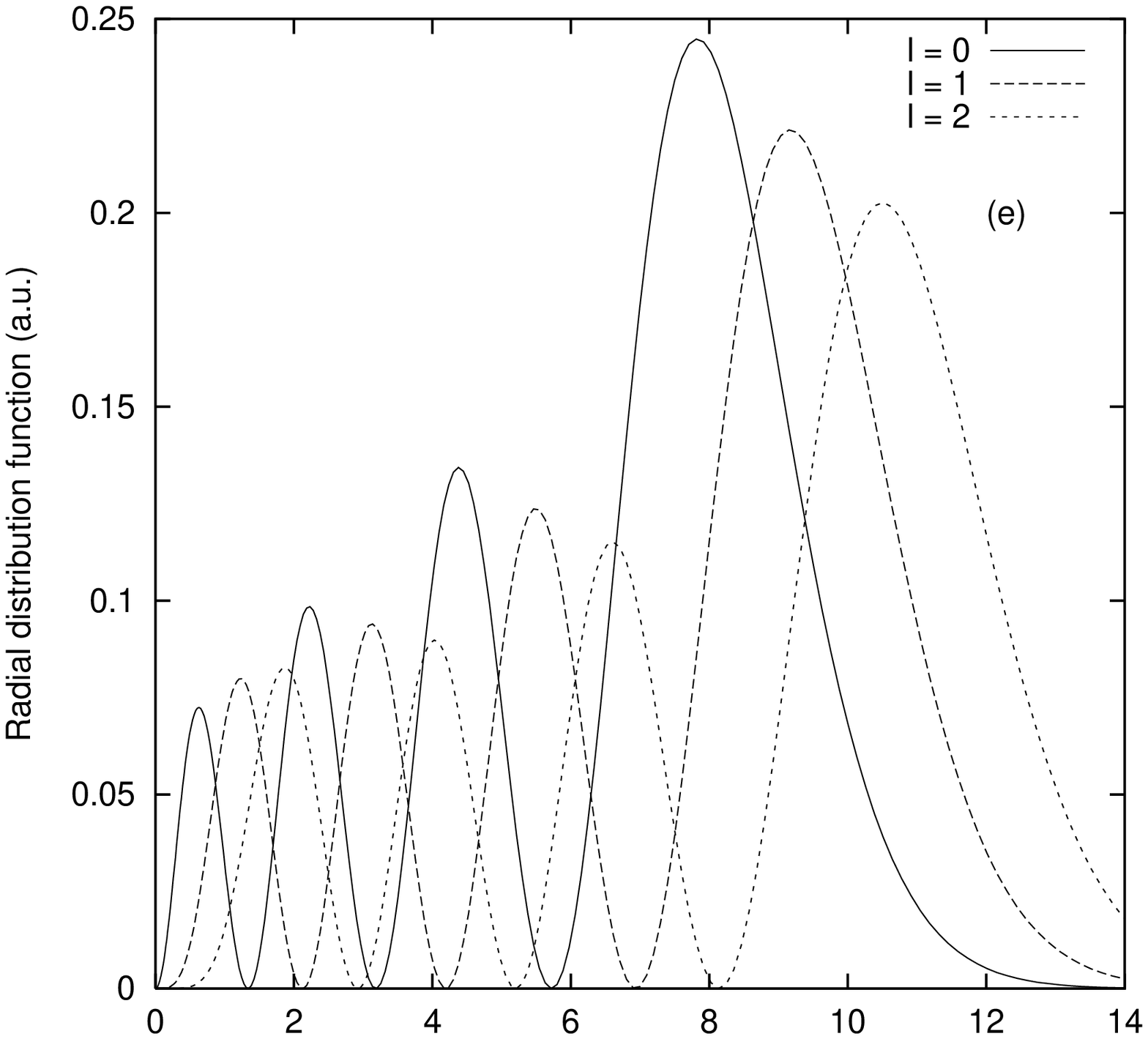}
\end{minipage}%
\\
\begin{minipage}[b]{0.40\textwidth}
\centering
\includegraphics[scale=0.27]{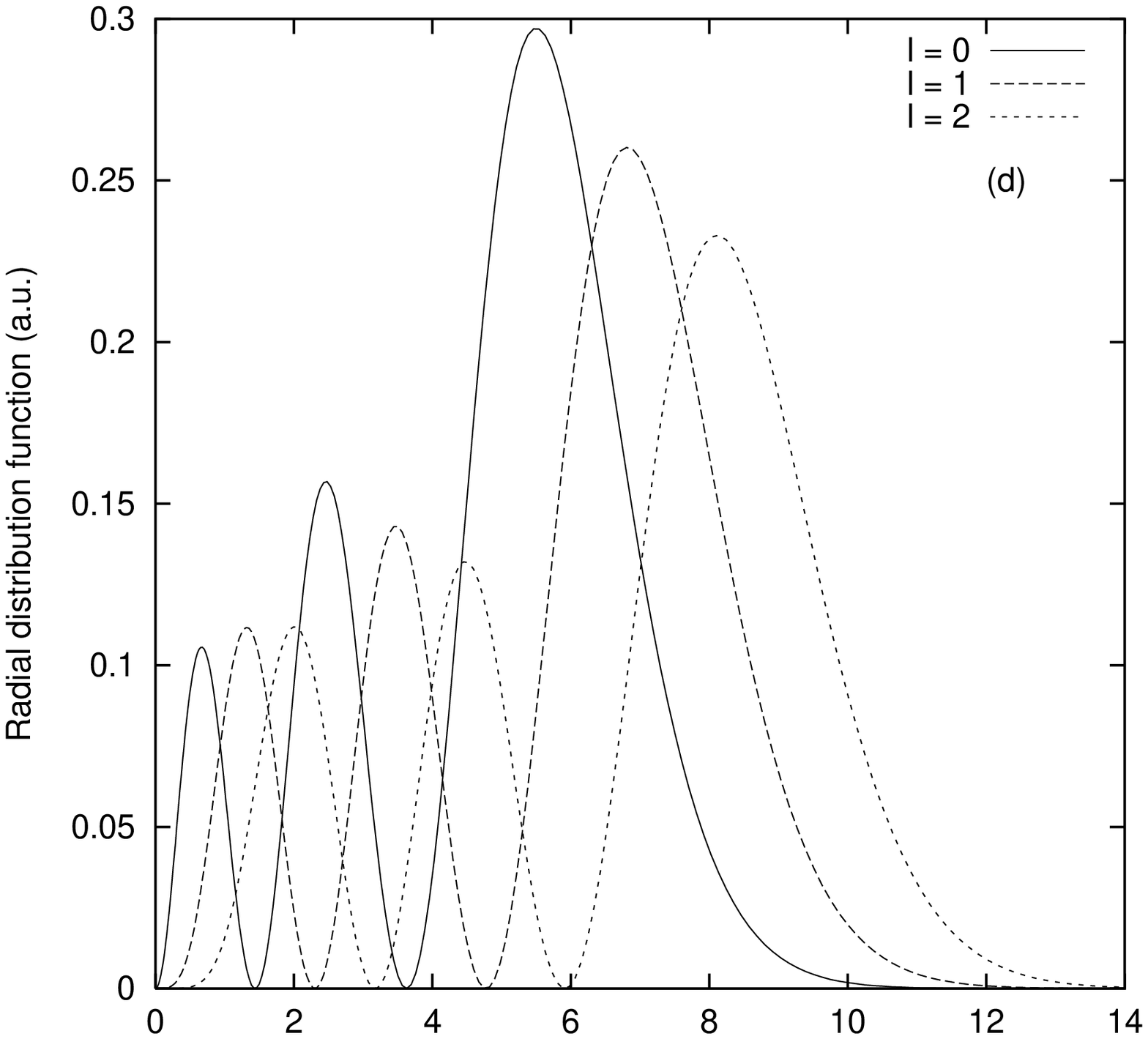}
\end{minipage}%
\\
\begin{minipage}[b]{0.40\textwidth}
\centering
\includegraphics[scale=0.27]{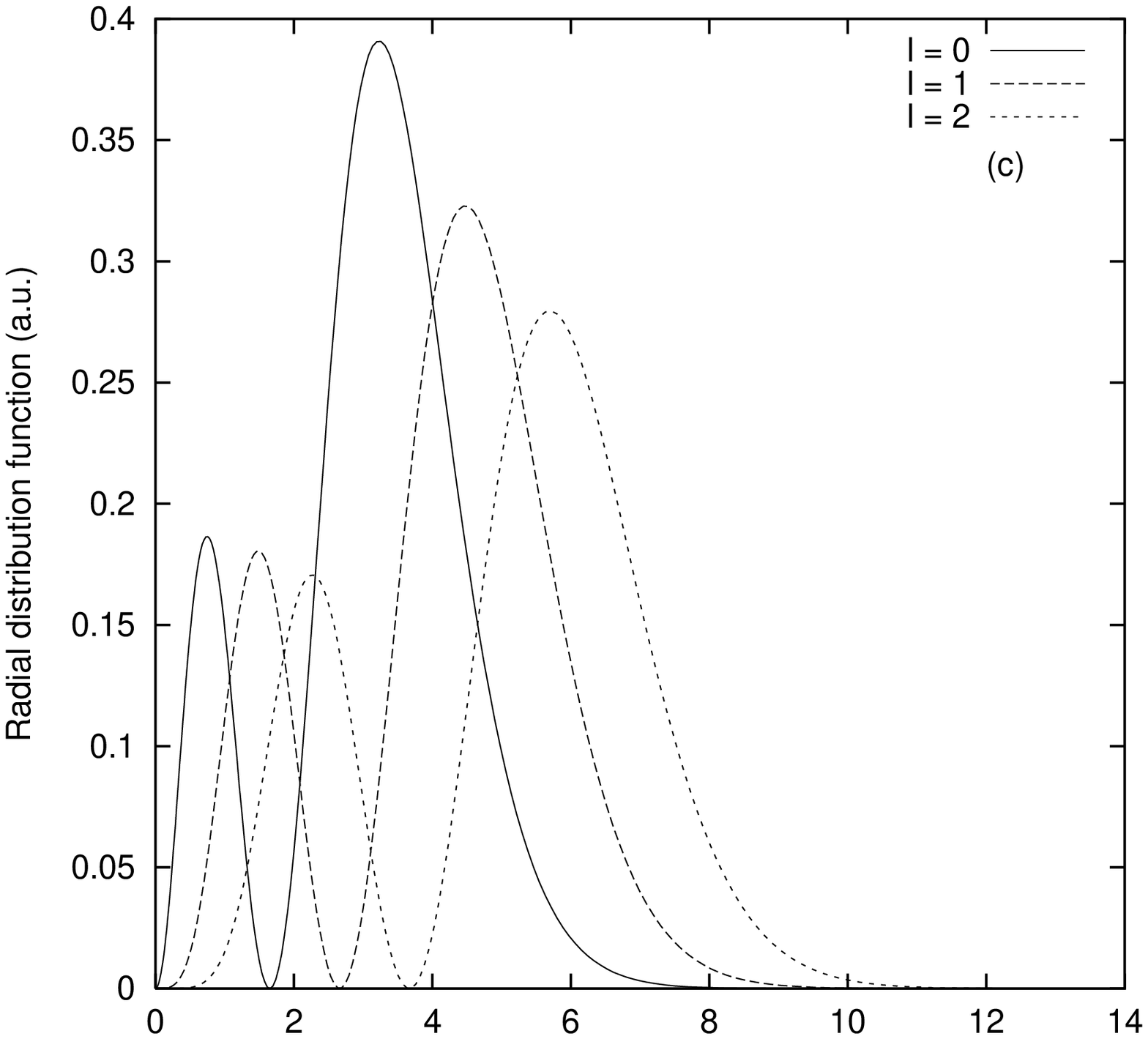}
\end{minipage}%
\\
\begin{minipage}[b]{0.40\textwidth}
\centering
\includegraphics[scale=0.27]{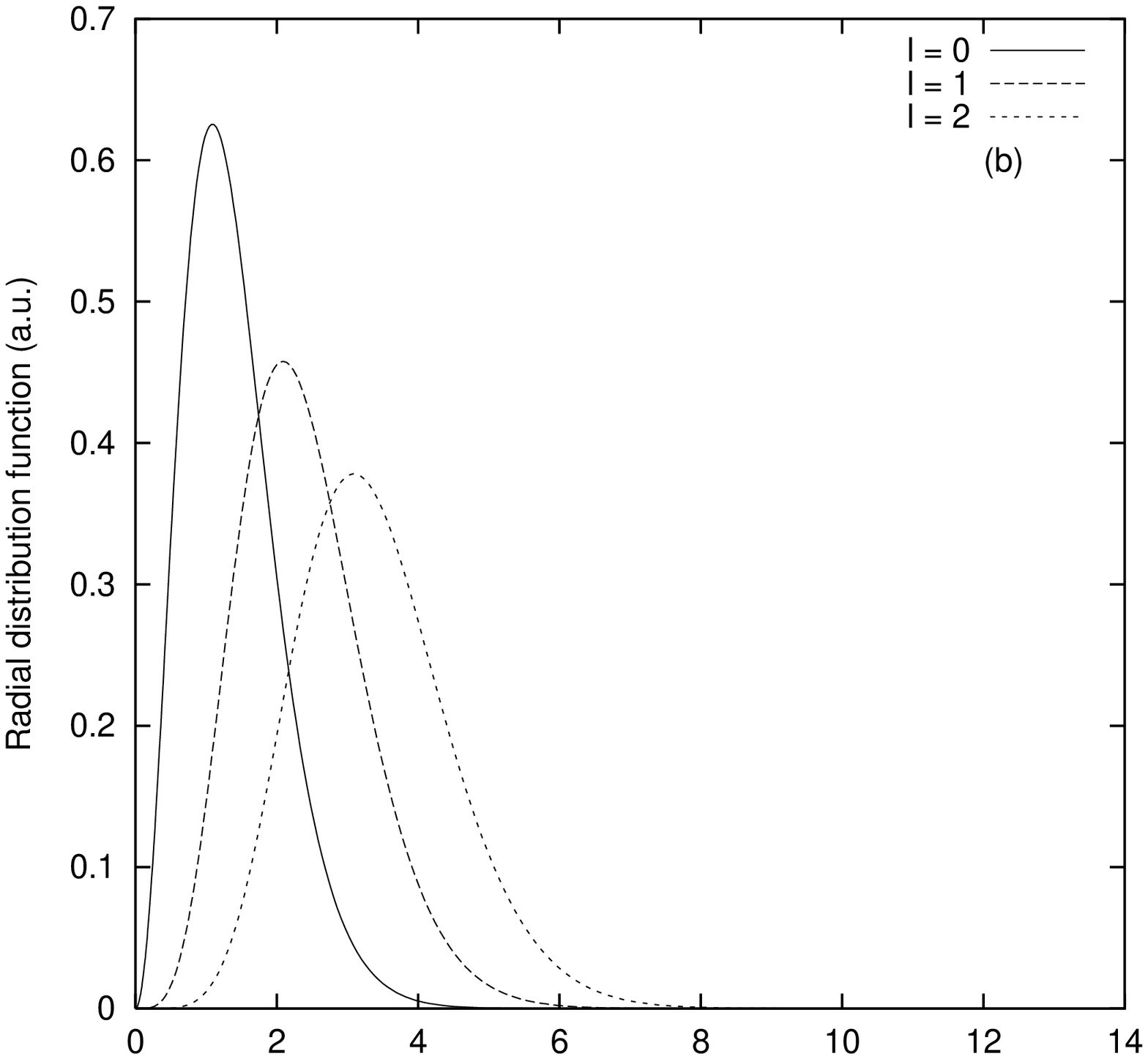}
\end{minipage}%
\\
\begin{minipage}[b]{0.40\textwidth}
\centering
\includegraphics[scale=0.27]{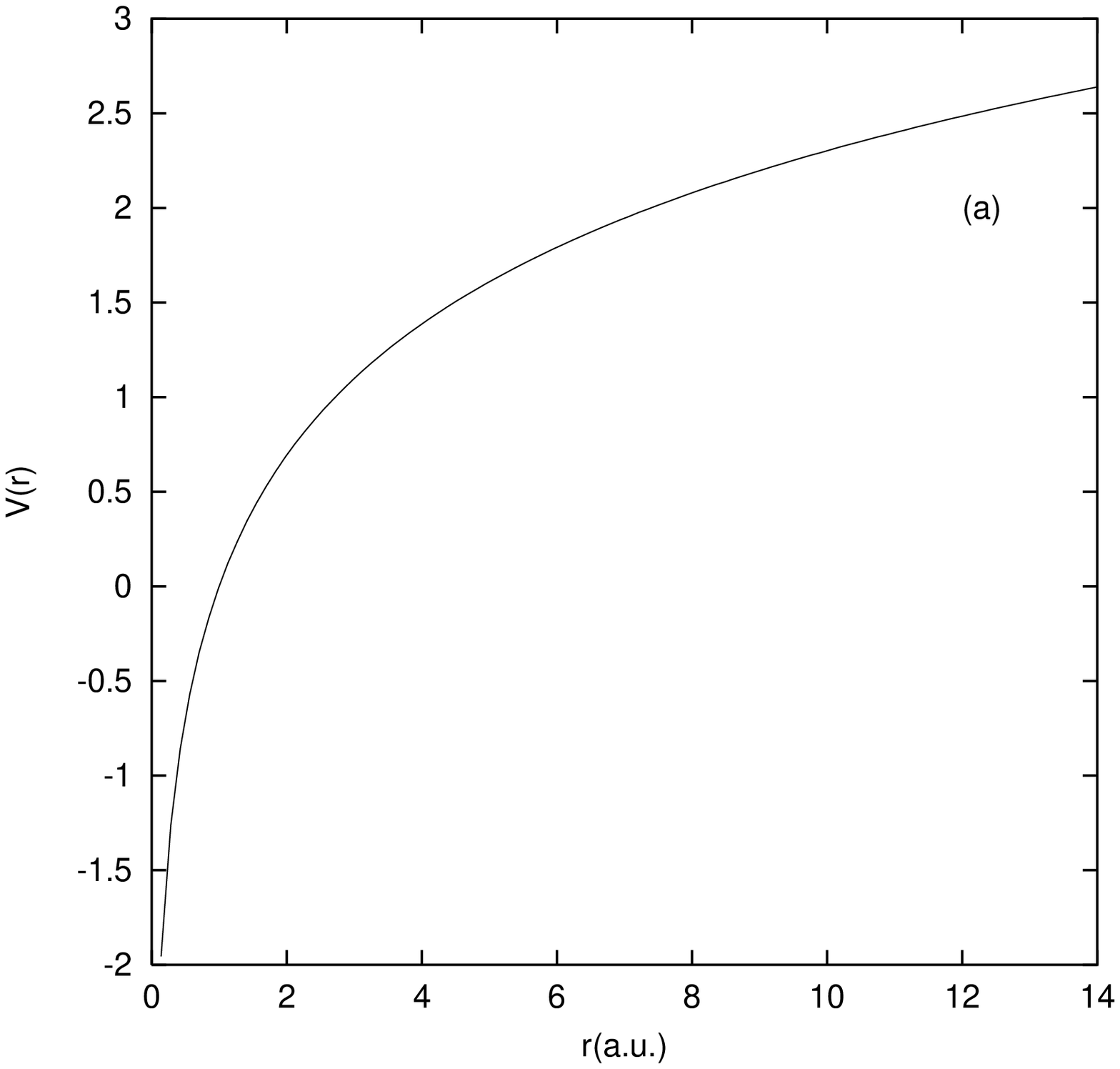}
\end{minipage}%

\caption{The radial probability distribution function, $|rR_{n\ell}|^2$ for 
the first four states corresponding to $\ell=0,1,2$  for the logarithmic 
potential. (a) the potential (b) ground state (c) first excited state (d) second excited 
state and (e) third excited state.}\label{fig:fig1}
\end{figure}

\begin{figure}
\begin{minipage}[b]{0.40\textwidth}
\centering
\includegraphics[scale=0.27]{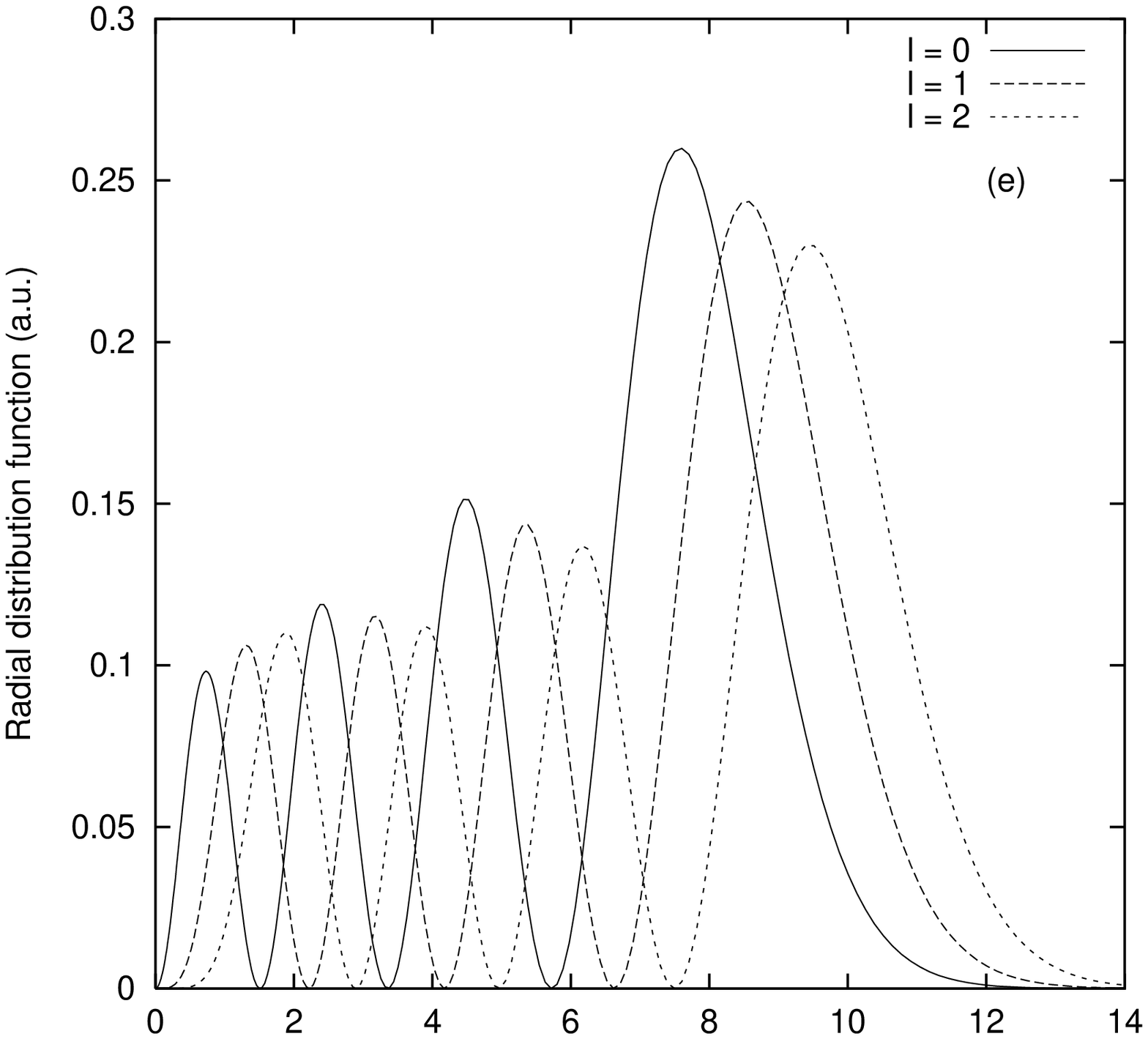}
\end{minipage}%
\\
\begin{minipage}[b]{0.40\textwidth}
\centering
\includegraphics[scale=0.27]{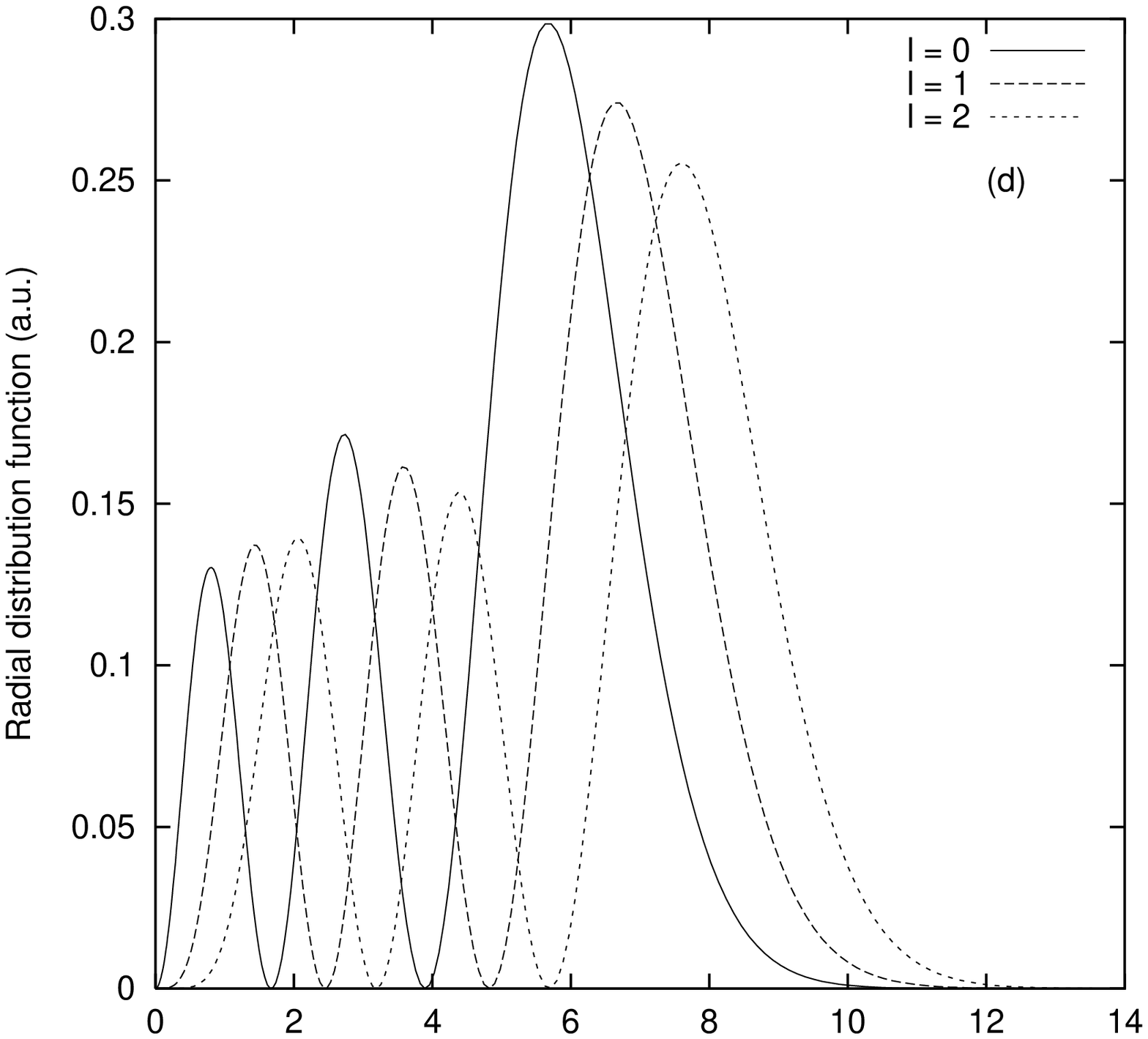}
\end{minipage}%
\\
\begin{minipage}[b]{0.40\textwidth}
\centering
\includegraphics[scale=0.27]{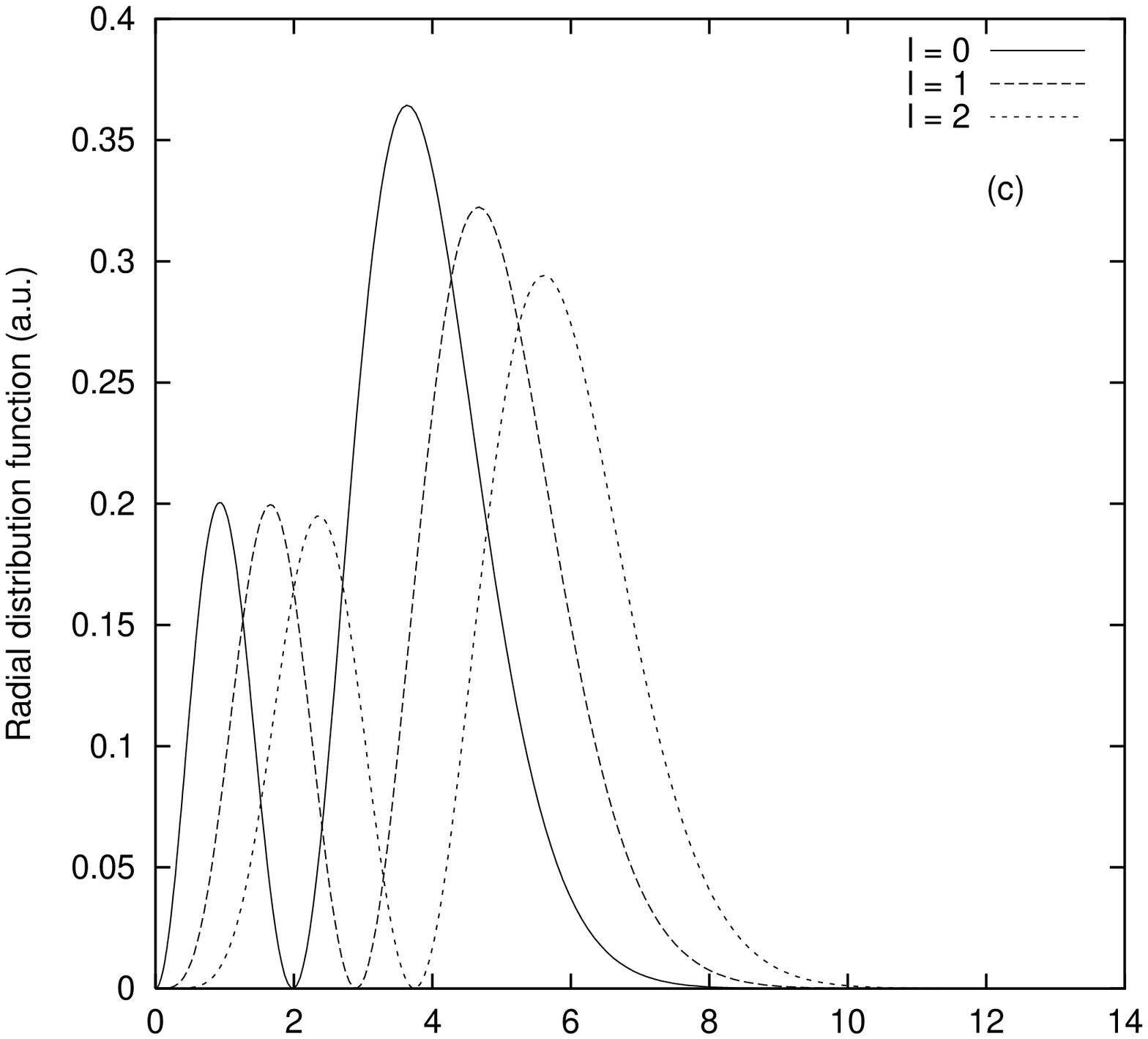}
\end{minipage}%
\\
\begin{minipage}[b]{0.40\textwidth}
\centering
\includegraphics[scale=0.27]{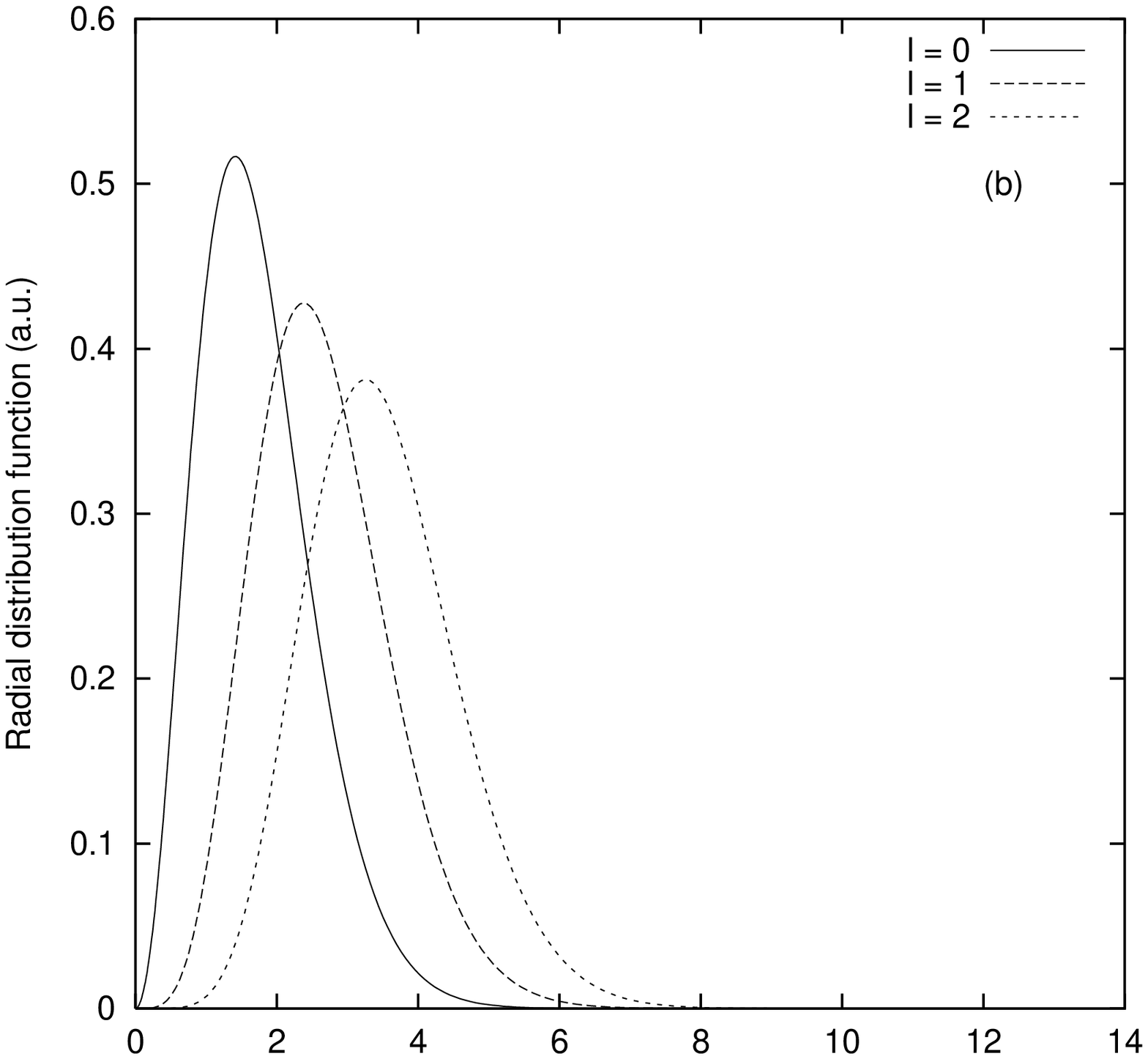}
\end{minipage}%
\\
\begin{minipage}[b]{0.40\textwidth}
\centering
\includegraphics[scale=0.27]{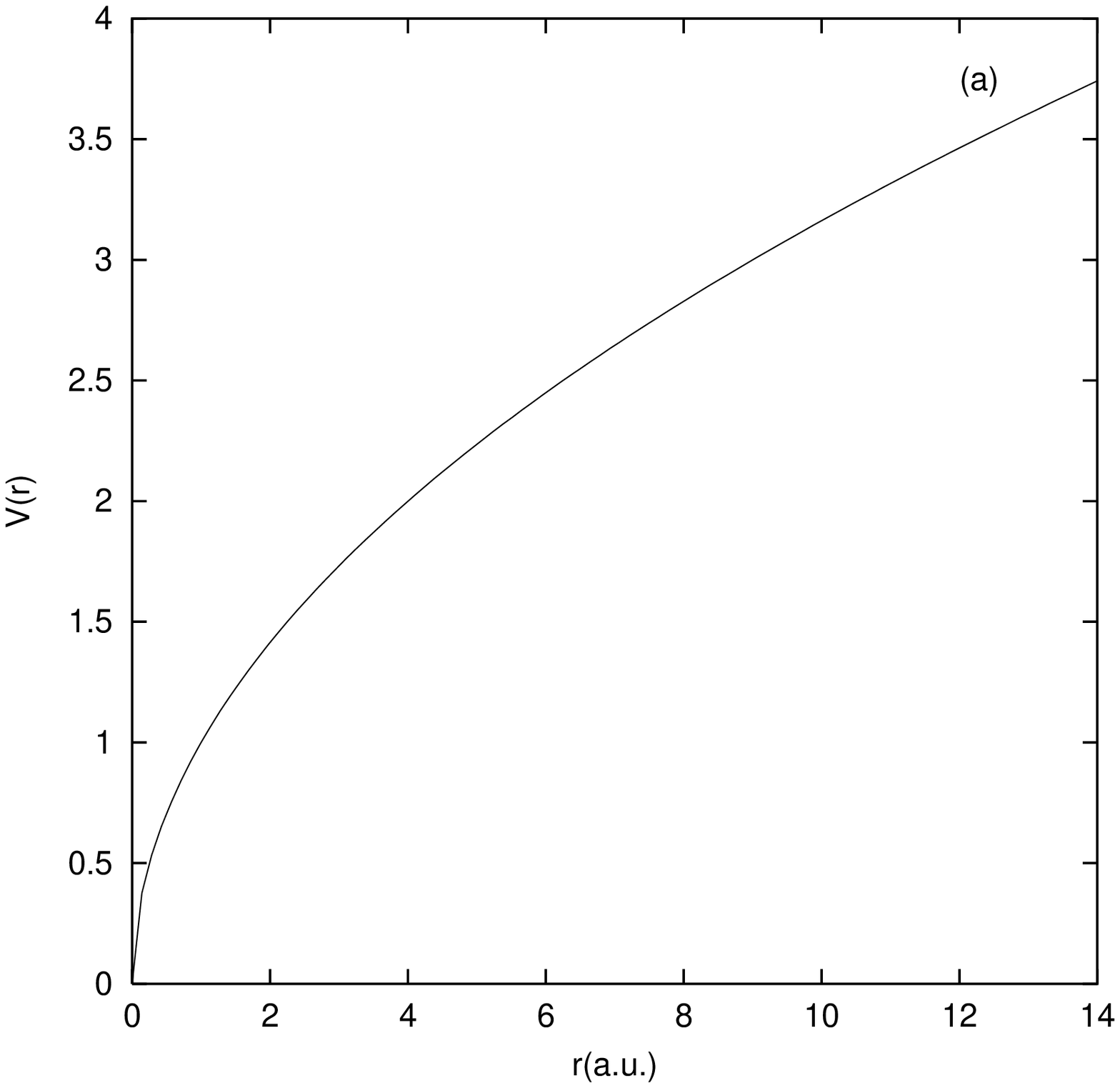}
\end{minipage}%

\caption{The radial probability distribution function, $|rR_{n\ell}|^2$ for 
the first four states corresponding to $\ell=0,1,2$ for the power law potential
$r^{0.5}$. (a) the potential (b) ground state (c) first excited state (d) second excited 
state and (e) third excited state.}\label{fig:fig2}
\end{figure}

\section{Conclusion}
The generalized pseudospectral method is shown to deliver arbitrary bound states 
of the power-law and logarithmic potentials. The method is simple, computationally
efficient, reliable and accurate. Application of this method on these systems show
that the current scheme offers results which are considerably better than the 
existing results available in the literature. It has the capability to handle the low
as well as the very high excited states with equal ease and accuracy which often pose 
problems for the variational
methods. As a test of the quality of the wave functions, we have also calculated 
the expectation values and the densities. 

\begin{acknowledgments}
I gratefully acknowledge the University of New Brunswick, Fredericton, New Brunswick,
Canada, for providing warm hospitality.
\end{acknowledgments}

\end{document}